\documentclass[times,review,]{elsarticle}

\usepackage{framed,multirow}
\usepackage[margin=1.2in]{geometry}
\usepackage{amssymb}
\usepackage{latexsym}
\usepackage{subfigure}
\usepackage{graphicx}

\usepackage{amsmath}

\DeclareMathOperator*{\argmin}{arg\,min}
\usepackage{booktabs}

\usepackage{url}
\usepackage{xcolor}
\usepackage{multirow}
\usepackage{hyperref}

\definecolor{newcolor}{rgb}{.8,.349,.1}

\journal{Computer Methods and Programs in Biomedicine}

\begin{document}


\begin{frontmatter}

\title{MA-RECON: Mask-aware deep-neural-network for robust fast MRI k-space interpolation  
}

\author[inst1]{Nitzan Avidan}
\affiliation[inst1]{organization={Faculty of Biomedical Engineering, Technion IIT},
            city={Haifa},
            country={Israel}}

\author[inst1]{Moti Freiman}





\begin{abstract}
{\bf Background and Objective:} High-quality reconstruction of MRI images from under-sampled `k-space' data, which is in the Fourier domain, is crucial for shortening MRI acquisition times and ensuring superior temporal resolution. Over recent years, a wealth of deep neural network (DNN) methods have emerged, aiming to tackle the complex, ill-posed inverse problem linked to this process. However, their instability against variations in the acquisition process and anatomical distribution exposes a deficiency in the generalization of relevant physical models within these DNN architectures. The goal of our work is to enhance the generalization capabilities of DNN methods for k-space interpolation by introducing `MA-RECON', an innovative mask-aware DNN architecture and associated training method. \\ 
{\bf Methods:} Unlike preceding approaches, our `MA-RECON' architecture encodes not only the observed data but also the under-sampling mask within the model structure. It implements a tailored training approach that leverages data generated with a variety of under-sampling masks to stimulate the model's generalization of the under-sampled MRI reconstruction problem. Therefore, effectively represents the associated inverse problem, akin to the classical compressed sensing approach. \\
{\bf Results:} The benefits of our MA-RECON approach were affirmed through rigorous testing with the widely accessible fastMRI dataset. Compared to standard DNN methods and DNNs trained with under-sampling mask augmentation, our approach demonstrated superior generalization capabilities. This resulted in a considerable improvement in robustness against variations in both the acquisition process and anatomical distribution, especially in regions with pathology.\\
{\bf Conclusion:} 
In conclusion, our mask-aware strategy holds promise for enhancing the generalization capacity and robustness of DNN-based methodologies for MRI reconstruction from undersampled k-space data.\\
Code is available in the following link: \url{https://github.com/nitzanavidan/PD_Recon}

\end{abstract}


\end{frontmatter}

\section{Introduction}
\label{sec:intro}
Magnetic resonance imaging (MRI), a non-invasive imaging modality, has broad clinical applications owing to its ability to yield detailed soft-tissue images~\cite{principles}. The signal for an MRI is gathered in the Fourier space, referred to as `k-space'. By applying an inverse Fourier transform (IFT) to the k-space, a meaningful MRI scan in the spatial domain is then generated. However, the requirement for extensive acquisition times to fully sample the k-space presents a significant barrier. Overcoming this obstacle is crucial for achieving high spatial and temporal resolutions, minimizing motion artifacts, enhancing patient comfort, and reducing costs~\cite{fastmri}.

Linear reductions in acquisition times can be achieved through partial sampling of the k-space. However, this approach presents a significant challenge: the reconstruction of an MRI image from under-sampled k-space data culminates in a highly ill-posed inverse problem. A Na\"\i ve approach to reconstruction, involving zero-filling the missing k-space data and subsequent application of the IFT, results in an image riddled with various artifacts, rendering it clinically meaningless~\cite{fastmri}. Initial efforts primarily focused on examining the properties of the k-space. For example, partial Fourier imaging methods exploited Hermitian symmetry to cut down acquisition times~\cite{feinberg1986halving}.

Traditional linear methods for MRI reconstruction from under-sampled data have capitalized on advancements in parallel imaging. This typically involves the use of multiple receiver coils in conjunction with linear algorithms for reconstruction. These can be applied either in the k-space domain~\cite{grappa} or in the spatial domain~\cite{sense}. However, the theoretical acceleration factor is inherently constrained by the number of accessible coils~\cite{hammernik2022physics}. Practical acceleration is further hindered due to noise amplification resulting from the process of matrix inversion~\cite{sense}.

The non-linear compressed sensing (CS) methodology strives to generate a high-quality image from under-sampled k-space data. It accomplishes this by constraining the accompanying ill-posed inverse problem with a sparsity prior via a sparsifying linear transform~\cite{cs}. Although the CS objective function lacks a closed-form solution, the issue remains convex. This allows it to be solved using a range of iterative algorithms~\cite{hammernik2022physics}. The SPIRiT technique enhances this process by further utilizing a non-linear regularized inverse problem formulation. This enables an auto-calibration-based reconstruction that is compatible with arbitrary sampling patterns~\cite{lustig2010spirit}.

In recent years, a multitude of deep-neural-network (DNN) based methods have been introduced for under-sampled MRI reconstruction. These methods have showcased considerable improvements in both image quality and acceleration factors compared to their traditional counterparts~\cite{chen2022ai,pawar2021domain,hong2023dual,wang2023mhan, wu2023deep,geng2023hfist}. Similar to classical techniques, DNN-based methods can be employed in both spatial and k-space domains.
One such model, the robust artificial neural networks for k-space interpolation (RAKI)~\cite{akccakaya2019scan}, utilizes a convolution neural network (CNN) to forecast interpolation kernels from a fully-sampled region of the k-space, mirroring the classical GRAPPA approach~\cite{grappa}.
DeepSPIRiT is another method that leverages CNN models to iteratively interpolate the missing k-space data on increasing areas of the k-space~\cite{cheng2018deepspirit}.
There are also dual-domain techniques such as KIKI-net~\cite{kiki} and DuDReTLU-net~\cite{hong2023dual}, which alternate iteratively between the image domain (I-CNN) and k-space (K-CNN). They incorporate a data consistency constraint to motivate the model to predict a physically plausible k-space interpolation.
Among more recent developments is the End-to-End Variational Network (E2E-VarNet). This method estimates coil-specific sensitivity maps and predicts the fully-sampled k-space from the under-sampled k-space data using a series of cascades. Each cascade merges a data-consistency module, operating in the k-space domain, with a refinement module functioning in the image domain~\cite{e2evarnet}. 

While the currently available DNN methods show encouraging results, their stability has been called into question. As demonstrated by Antun et al. \cite{instabilities} and Jalal et al. \cite{jalal2021robust}, unlike classical methods, DNN-based approaches are susceptible to instability in the face of variations in the acquisition process and anatomical distribution. Such variations might arise from the use of different under-sampling masks or acceleration factors during inference compared to those utilized during training. Likewise, instability can be introduced by the presence of minor pathologies or differing anatomies compared to the data used for training.

Initial efforts to bridge the stability gap in DNN-based MRI reconstruction focused on data augmentation techniques. Specifically, Liu et al.~\cite{santis} enhanced the overall reconstruction performance and resilience against discrepancies in sampling patterns and images acquired at varying contrast phases. This was achieved by enriching the under-sampled data with a wide array of under-sampling patterns. More recently, Jalal et al.~\cite{jalal2021robust} employed a hybrid approach, combining DNN-based generative priors with classical CS-based reconstruction and posterior sampling, to address the stability gap. However, it's worth noting that this method doesn't directly tackle the inherent stability gap present in DNN-based MRI reconstruction techniques.

Recently, physics-driven deep learning methods have emerged as potent tools for enhancing the generalization capacity of DNN-based under-sampled MRI reconstruction. These encompass methods that incorporate the physics of MRI acquisitions through physics-driven loss functions~\cite{yang2017dagan,hyun2018deep,cui2022,hong2023dual,zhang2021dual}, plug-and-play methods~\cite{yazdanpanah2019deep,ahmad2020plug}, generative models~\cite{quan2018compressed,shaul2020subsampled,Zach2022,zhao2023swingan,lv2021transfer}, and unrolled networks~\cite{liang2019deep,hammernik2022physics}.

However, the stability of DNN-based methods in the face of variations in the acquisition process and anatomical distribution still remains a crucial open question~\cite{hammernik2022physics}.

\begin{figure*}[h!]

    \centering
    \subfigure[]{\includegraphics[width=0.17\textwidth]{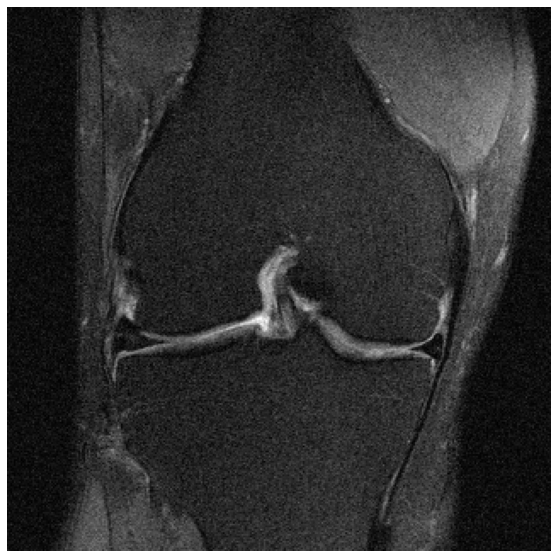}}
    \hfill
    \subfigure[]{\includegraphics[width=0.17\textwidth]{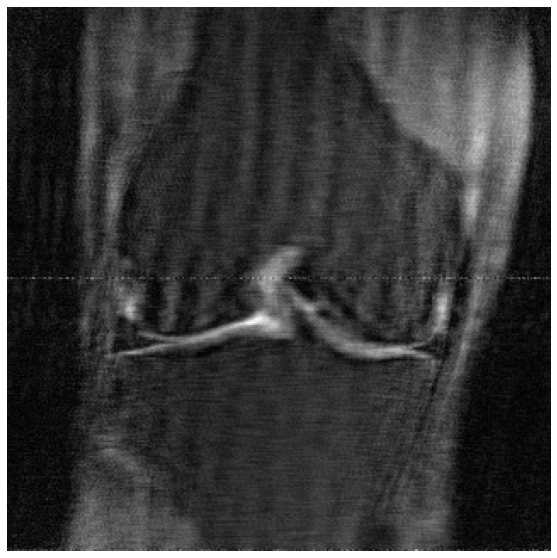}}
    \hfill
    \subfigure[]{\includegraphics[width=0.17\textwidth]{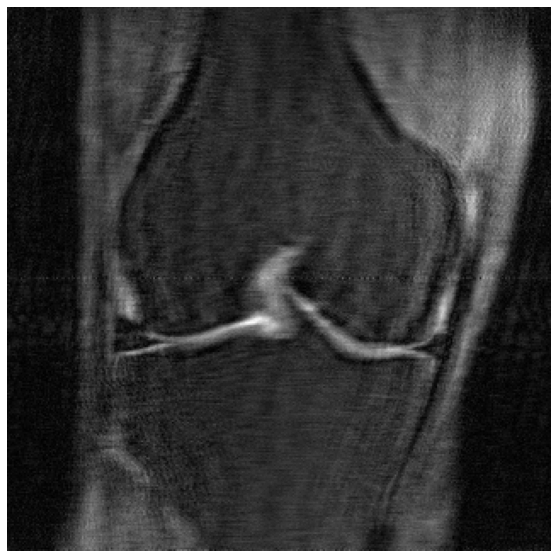}}
    \hfill
    \subfigure[]{\includegraphics[width=0.17\textwidth]{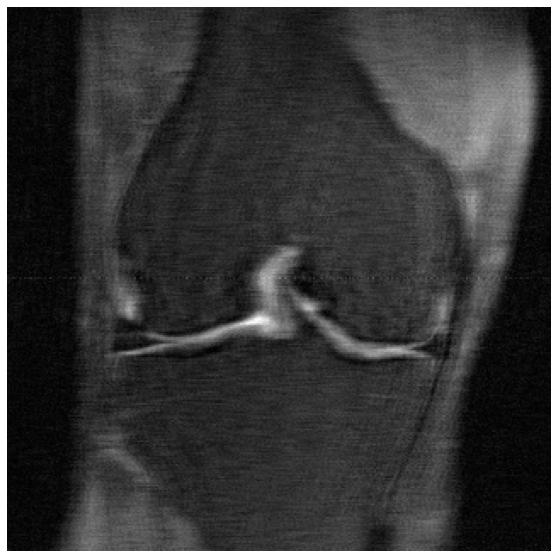}}
    
    \subfigure[]{\includegraphics[width=0.17\textwidth]{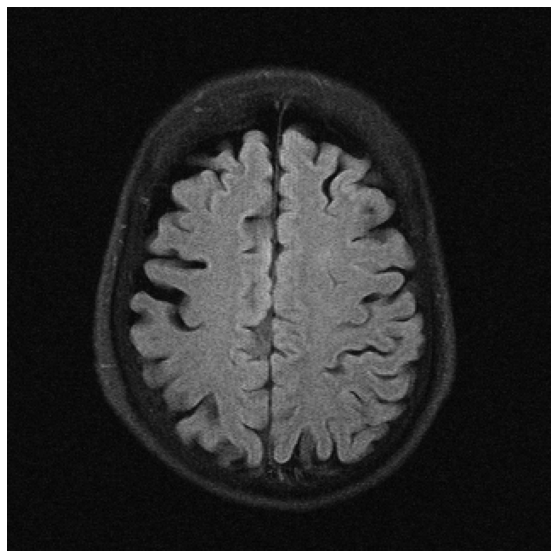}}
    \hfill
    \subfigure[]{\includegraphics[width=0.17\textwidth]{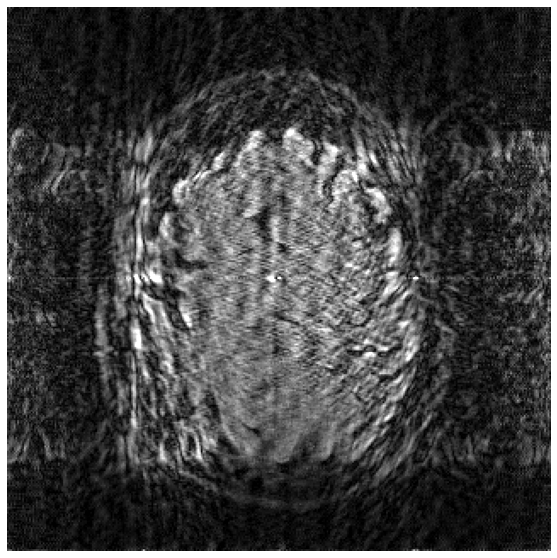}}
    \hfill
    \subfigure[]{\includegraphics[width=0.17\textwidth]{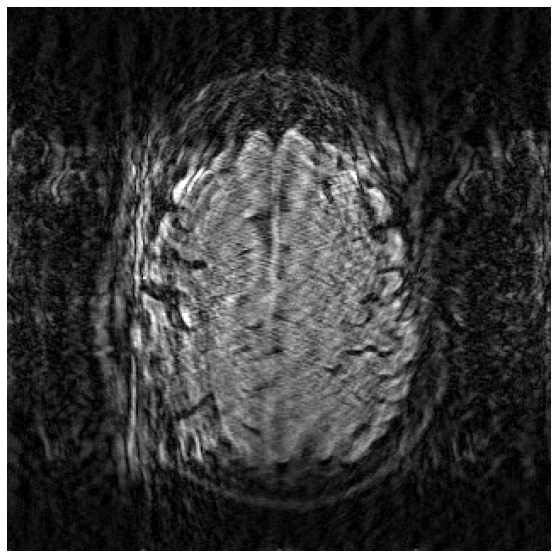}}
    \hfill
    \subfigure[]{\includegraphics[width=0.17\textwidth]{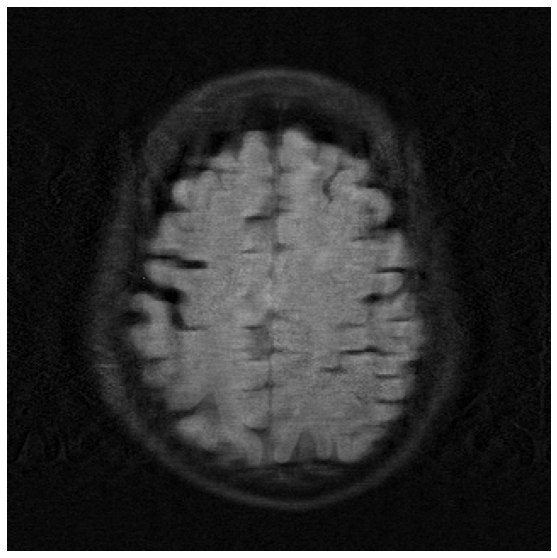}}
    
    \caption{The stability gap in DNN-based MRI reconstruction from undersampled data. The First row depicts variation in the acquisition process (different undersampling masks) while the second row presents a variation in the anatomical distribution (train on knee data and inference on brain data). (a/e) The target image from fully-sampled k-space data, (b/f) reconstruction using a model trained with a fixed sampling mask (PSNR=24.9/14.07, SSIM=0.4684/0.1926), (c/g) reconstruction using a model trained with mask augmentations (PSNR=25.35/16.56, SSIM=0.4969/0.2812), and; (d/h) reconstruction using the proposed MA-RECON approach ({\bf PSNR=26.4/26.05, SSIM=0.5354/0.6359)}. }
    
    \label{fig:intro}
    
\end{figure*}

In this work, our objective is to bridge this stability gap in under-sampled MRI reconstruction with DNN by presenting MA-RECON - a novel mask-aware DNN architecture and training methodology. Unlike preceding approaches, our architecture encodes not only the observed data but also the under-sampling mask within the model structure. It implements a tailored training approach that leverages data generated with a variety of under-sampling masks to stimulate the model's generalization of the under-sampled MRI reconstruction problem. Our model, by encoding the under-sampling mask in addition to the observed data in the model architecture, effectively represents the associated inverse problem, akin to the classical CS approach. This approach incorporates the sampling mask as part of the system's forward model during the optimization process. Fig.~\ref{fig:intro} succinctly demonstrates our primary result, which illustrates the improved generalization capacity and resilience against variations - both in the acquisition process (i.e., differences in the under-sampling masks) and in the anatomical distribution (i.e., training on knee data and inference on brain data) - of our MA-RECON approach (d,h) compared to the baseline methods (b,f, and c,g).

Our distinct contributions in this work can be outlined as follows: 1) the introduction of a mask-aware DNN methodology for k-space interpolation from under-sampled 'k-space' data, 2) the enhancement of generalization capacity and resilience against variations in both the acquisition process and anatomical distribution, and; 3) the provision of experimental evidence substantiating improved robustness, particularly in clinically pertinent regions, utilizing the publicly accessible fastMRI dataset~\cite{fastmri,zhao2021fastmri+}.

This manuscript provides a comprehensive exploration of our initial research that was accepted for presentation at the  2023 MICCAI Workshop on Clinical image-based procedures (CLIP). The current work offers an expanded elucidation of the methodology and delves into the generalization capacity of the technique. To accomplish this, we executed supplementary experiments incorporating both knee and brain data to evaluate the generalization efficacy of the approach when faced with shifts in anatomical distribution. 

\section{Background}
\label{sec:background}
The forward model of the undersampled MRI acquisition process is given by:
\begin{equation}
    \label{eq:forward_model}
    k_{us} = M \circ \mathcal{F}x + n
\end{equation}
where $k_{us} \in \mathbb{C}^N$ are the observed measurements in the k-space, $x\in \mathbb{R}^N$ is the image representing the underlying anatomy, $\mathcal{F}$ is the Fourier operator, $M \in \mathbb{R}^N$ is a binary undersampling mask, $\circ$ is element-wise multiplication and $n$ is an additive noise. For the sake of simplicity, we assume $n \sim \mathcal{N}\left(0,\sigma^2\right)$ \cite{aja2016statistical}. 

Direct reconstruction of the MRI image $x$ from the undersampled data is an ill-posed inverse problem that cannot be simply solved using linear approaches. Na\"ive reconstruction by zero-filling of the missing k-space data and application of the IFT will result in an aliased image which is clinically meaningless~\cite{fastmri}.  

The non-linear CS approach enables high-quality reconstruction from the undersampled k-space by imposing a sparsity constraint, $\Psi$ to regularize the ill-posed inverse problem. The reconstructed image, $\hat{x}$, is obtained by solving the following constrained optimization problem: 
\begin{equation}
    \label{eq:cs}
    \hat{x} = \argmin_{x}  \left\|M \circ \mathcal{F} (x) - k_{us}\right\|_2+ \lambda\Psi(x)
\end{equation}
Where $\lambda$ is the regularization weight balancing between the data term, and the assumed prior. Examples of the sparsifying transform $\Psi$ include the total-variation and the wavelet transforms~\cite{cs}. Recently, Jalal et al.~\cite{jalal2021robust} suggested to replace the sparsifying transform $\Psi$ with a DNN-based generative prior. Various optimization techniques were developed to address the challenging CS optimization problem~\cite{ye2019compressed}. Yet, the high computational complexity and limited ability to overcome image quality degradation at high acceleration rates may interfere with clinical utilization~\cite{ye2019compressed}. 

Recently, DNN-based methods were applied for MRI reconstruction from undersampled data. In the k-space domain, these methods aim to predict the fully sampled k-space data from the given undersampled k-space data. Without loss of generality, the prediction task can be represented as:
\begin{equation}
    \label{eq:baseline_eq}
    \widehat{k_{full}} = F_{\theta}\left(k_{us}\right)
\end{equation}
where $F_{\theta}$ denotes the network function and $\theta$ represents the DNN weights.
Taking a supervised learning approach, the DNN weights $\theta$ are estimated by minimizing the loss between the full k-space data predicted by the DNN, $\widehat{k_{full}}$, from the undersampled data, $k_{us}$, to the corresponding ground truth fully sampled k-space data, $k_{full}$, as follows:
\begin{equation}
    \label{eq:baseline_train_eq}
    \hat{\theta} = \argmin_{\theta} \sum^{n_{data}}_{i=0}\left\|F_\theta\left(k_{us}^{(i)}\right) - k_{full}^{(i)}\right\|_1
\end{equation}
However, unlike the classical CS approach (Eq.~\ref{eq:cs}), such methods are known to be unstable against the presence of variations in the acquisition process and the anatomical distribution~\cite{instabilities,jalal2021robust}.

A key observation is that while the CS approach (Eq.~\ref{eq:cs}) explicitly encodes the undersampling mask $M$ as part of the forward model, DNN-based approaches essentially ignore the undersampling mask during training and inference. Even though augmentation techniques suggest to use undersampled k-space data from different masks during training~\cite{santis}, or in physically-motivated loss functions~\cite{hammernik2022physics}, the undersampling mask information is not explicitly encoded in the DNN architecture nor leveraged during inference. This may result in DNN instability compared to the classical CS counterpart. 
\vspace{1cm}

\begin{figure*}[th!]
     \centering
     \includegraphics[width=0.95\textwidth]{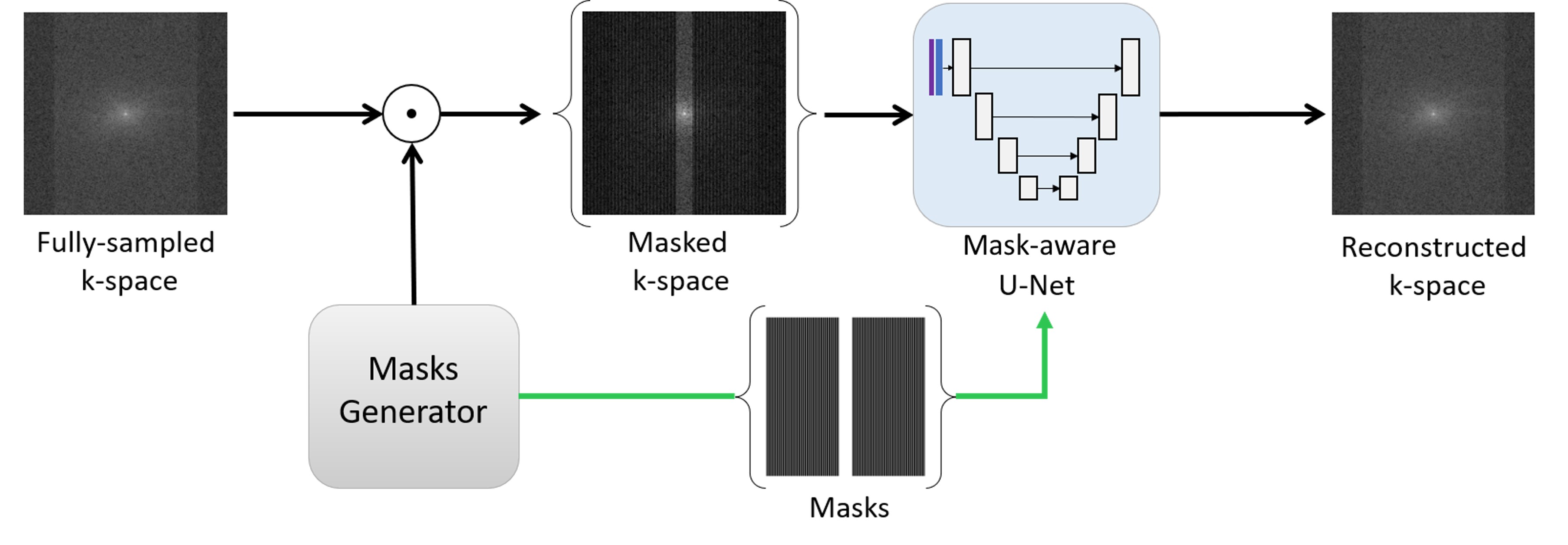}
     \caption{The fully sampled k-space is multiplied by a binary mask to create the undersampled k-space. Then, the mask (purple line) used for the sampling is fed into the mask-aware DNN model in addition to the undersampled k-space (blue line). }
     \label{fig:scheme}
\end{figure*}

\section{Methods}
\label{sec:methods}
\subsection{Mask-aware DNN for k-space interpolation}
Our main hypothesis is that by explicitly encoding the undersampling mask in the DNN architecture and leveraging this information during inference, the DNN will be more capable of accurately generalizing the ill-posed inverse problem associated with MRI reconstruction from undersampled data. Thus, it will be more robust against variations in the acquisition process and anatomical distribution.  
We introduce a mask-aware, U-Net~\cite{unet} based, DNN architecture operating on the k-space domain. We simplify the multi-coil problem by combining the images obtained from each coil into a single image, such that we essentially solve a single-coil problem. Although our dataset contained multi-coil knee imaging data, it lacked the essential sensitivity maps required for the reconstruction network. We opted to work with a single-coil dataset to isolate the mask's impact in the reconstruction process. This approach allowed us to distinctly assess the mask's contribution to the network's performance, avoiding the complexities introduced by multi-coil data lacking sensitivity maps. The complex-valued k-space data is represented as a two-channel input, corresponding to the real and imaginary parts. The undersampling mask $M$ is encoded by adding a 3$^{\textrm{rd}}$ input channel to the DNN. 
The prediction of the full k-space data from the undersampled k-space data is defined as:
\begin{equation}
\label{eq:mask_eq}
   \widehat{k_{full}} = F_{\theta}\left(k_{us}, M\right)
\end{equation}
The DNN weights $\theta$ are estimated by minimizing the loss between the full k-space data predicted by the DNN, $\widehat{k_{full}}$, from the undersampled data, $k_{us}$ and the undersampling mask $M$, to the corresponding ground truth fully sampled k-space data, $k_{full}$, as follows:
\begin{equation}
\label{mask_train_eq}
   \hat{\theta} = \argmin_{\theta} \sum^{n_{data}}_{i=0}\left\|F_\theta\left(k_{us}^{(i)}, M^{(i)}\right) - k_{full}^{(i)}\right\|_1
\end{equation}
We encourage our mask-aware DNN model to generalize the ill-posed inverse problem of MRI reconstruction from undersampled k-space data by varying the undersampling mask $M$ during training. Fig.~\ref{fig:scheme} illustrates our mask-aware approach for training.

\subsection{Implementation details}
Our models are based on the suggested fastMRI U-Net~\cite{fastmri}, comprising two deep convolutional networks, an encoder followed by a decoder. In the MA-RECON network, the input to the encoder is a 3-channel data representing the concatenation of the complex k-space data (2 channels) and the binary undersampling mask (1 channel). In addition, we modified the output to be the summation of the k-space input and the network output to facilitate residual learning. The encoder consists of blocks of two 3$\times$3 convolutions, each followed by Rectified Linear Unit (ReLU) activation functions. The output of each block is down-sampled using a max-pooling layer with stride 2. The decoder consists of blocks with a similar structure to the encoder, where the output of each block is up-sampled using a bilinear up-sampling layer. The decoder concatenates the two inputs, the up-sampled output of the previous block and the output of the encoder block with the same resolution, to the first convolution in each block. At the end of the encoder, there is one 1$\times$1 convolution that reduces the number of channels to two channels, representing the real and imaginary parts of the k-space data. The detailed model architecture is provided as an appendix.  

\section{Experiments}
\label{experiments}
\subsection{Dataset}
We used the publicly available fastMRI dataset~\cite{fastmri}, consisting of raw k-space data of knee and brain volumes. The knee images used in this study were directly obtained from the single-coil track of the fastMRI dataset, while the brain images were reconstructed from fully sampled, multi-coil k-space data. To create a single-coil complex valued brain dataset, we applied the Inverse Fourier Transform (IFT) to each individual coil. We then combined the coil images using geometric averaging, which effectively served as a sensitivity-weighted combination. This approach yielded a satisfactory result.
The training set consisted of 34742 knee slices. We split the fastMRI knee validation set into validation and test sets since the original fastMRI test set does not allow applying random undersampling. The splitting ratio was 2:1, yielding 5054 slices for validation and 2081 slices for testing. Brain images (1000 slices) were used for test purposes only.

To further evaluate the clinical impact of our approach, we also used the bounding box annotations generated by subspecialist experts on the fastMRI knee and brain dataset provided by the fastMRI+ dataset~\cite{zhao2021fastmri+}. Each bounding box annotation includes its coordinates and the relevant label for a given pathology on a slice-by-slice level.
Pathology annotations were marked in 39\% of our training set, 28\% of the validation set, and 51\% of the test set. These annotations enabled us to examine the clinical relevance of our models' performances on pathological regions. 

\subsection{Undersampling masks}
\begin{figure}[t!]
     \centering
     \subfigure[]{\includegraphics[width=0.20\textwidth]{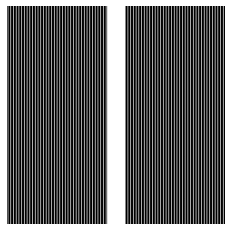}}
     \subfigure[]{\includegraphics[width=0.20\textwidth]{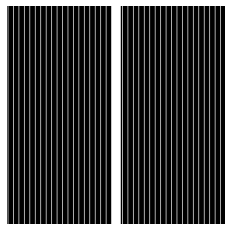}}
     \subfigure[]{\includegraphics[width=0.20\textwidth]{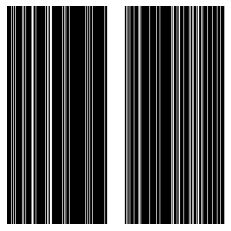}}
     \subfigure[]{\includegraphics[width=0.20\textwidth]{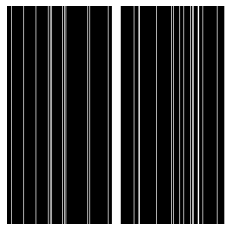}}
     
     \subfigure[]{\includegraphics[width=0.20\textwidth]{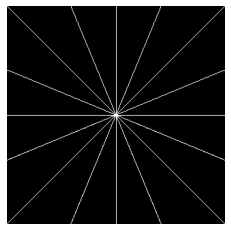}}
     \subfigure[]{\includegraphics[width=0.20\textwidth]{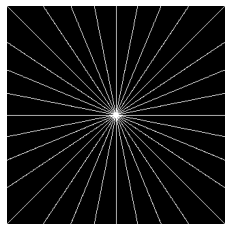}}
     \subfigure[]{\includegraphics[width=0.20\textwidth]{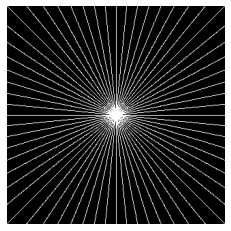}}
     \subfigure[]{\includegraphics[width=0.20\textwidth]{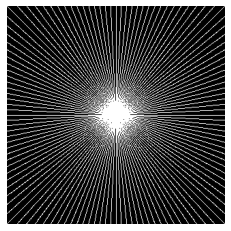}}

     \caption{Masks patterns. (a) equispaced mask with R=4, (b) equispaced mask with R=8, (c) random mask with R=4, (d) random mask with R=8, (e) radial mask with 8 spokes - R$\approx$40, (f) radial mask with 16 spokes - R$\approx$20,  (g) radial mask with 32 spokes - R$\approx$10, and;(h) radial mask with 64 spokes - R$\approx$5}
    \label{fig:masks}
\end{figure}

We retrospectively undersampled the fully sampled k-space data by element-wise multiplication with randomly generated binary masks. The acceleration factor (R) was set to four or eight, where the undersampled k-space included 8\% or 4\% of the central region, respectively. The remaining k-space lines were sampled in three different ways to achieve the desired acceleration factor: 1) an equispaced mask with a fixed offset; the remaining k-space lines were sampled with equal spacing, 2) an equispaced mask with a varying offset; the remaining k-space lines were sampled with equal spacing but with random offset from the start. The number of possible masks depends on the desired acceleration factor. For instance, if the acceleration factor is 3, three different masks would be generated. 3) a random mask, meaning that the remaining k-space lines were uniformly sampled. We also used radial sampling patterns consisting of 8, 16, and 32 spokes for inference. Figure~\ref{fig:masks} depicts representative examples of the undersampling masks.

\subsection{Training settings}
We trained three models on the undersampled k-space with R=4: 1) undersampled k-space with a fixed equispaced pattern (Fixed), 2) undersampled k-space with mask augmentations; varying equispaced pattern (Baseline), and; 3) undersampled k-space with mask augmentations, while the input consists of three channels where the third channel is the mask pattern (MA-RECON).
We trained the models on transformed k-space to account for the orders of magnitude difference between the DC coefficient and the rest of the coefficients in the k-space. Specifically, we used the following transformation on the real and imaginary part of the k-space :
\begin{equation}
k_{t} = \log(k + 1) \cdot 10^5
\end{equation}
All models were trained using the RMSprop algorithm. An initial learning rate of 0.01 was used for all models. The initial learning rate was multiplied by 0.1 after 40 epochs for all models. We used the L1 norm between the 2 channels fully sampled k-space and the 2 channels reconstructed k-space as the loss function. 
We trained our models on two Nvidia A100 GPUs, each for 200 epochs. Training time took about 48 hours for each model. We selected the models with the best validation loss for our experiments.

\subsection{Experimental methodology}
We examined the generalization capability of the networks by testing their performance with varying acquisition conditions and anatomical distributions. Specifically, we evaluated reconstruction performance for: 
\begin{enumerate}
    \item undersampling patterns different from those used during training; equispaced mask for training and random or radial masks for inference.
    \item different acceleration factors; R=4 for training and R=8 for inference.
    \item different anatomical distribution; knee for training and brain for inference.
\end{enumerate}
To specifically determine the clinical relevance of our approach, we calculated performance metrics separately for the entire images and for clinically relevant regions that include pathologies. We used the clinical regions annotations from the fastMRI+ dataset~\cite{zhao2021fastmri+}. 

The evaluation was performed using standard evaluation metrics, including Peak Signal-to-Noise Ratio (PSNR), and Structural Similarity (SSIM) to assess the different models' performances. 
We determined statistically significant differences between our approach and a baseline model trained with undersampling mask data augmentation with a paired Student's t-test. 

Finally, we qualitatively assessed the robustness against anatomical distribution shift of our MA-RECON in comparison to the state-of-the-art E2E VarNet reconstruction model~\cite{e2evarnet}. For that purpose, we used a pretrained E2E VarNet model trained on the knee dataset for inference on the brain dataset~\cite{fastmri}.

\section{Results}
\subsection{Variation in the acquisition process}
\begin{figure}[t!]
     \centering
     \subfigure[]{\includegraphics[width=0.45\textwidth]{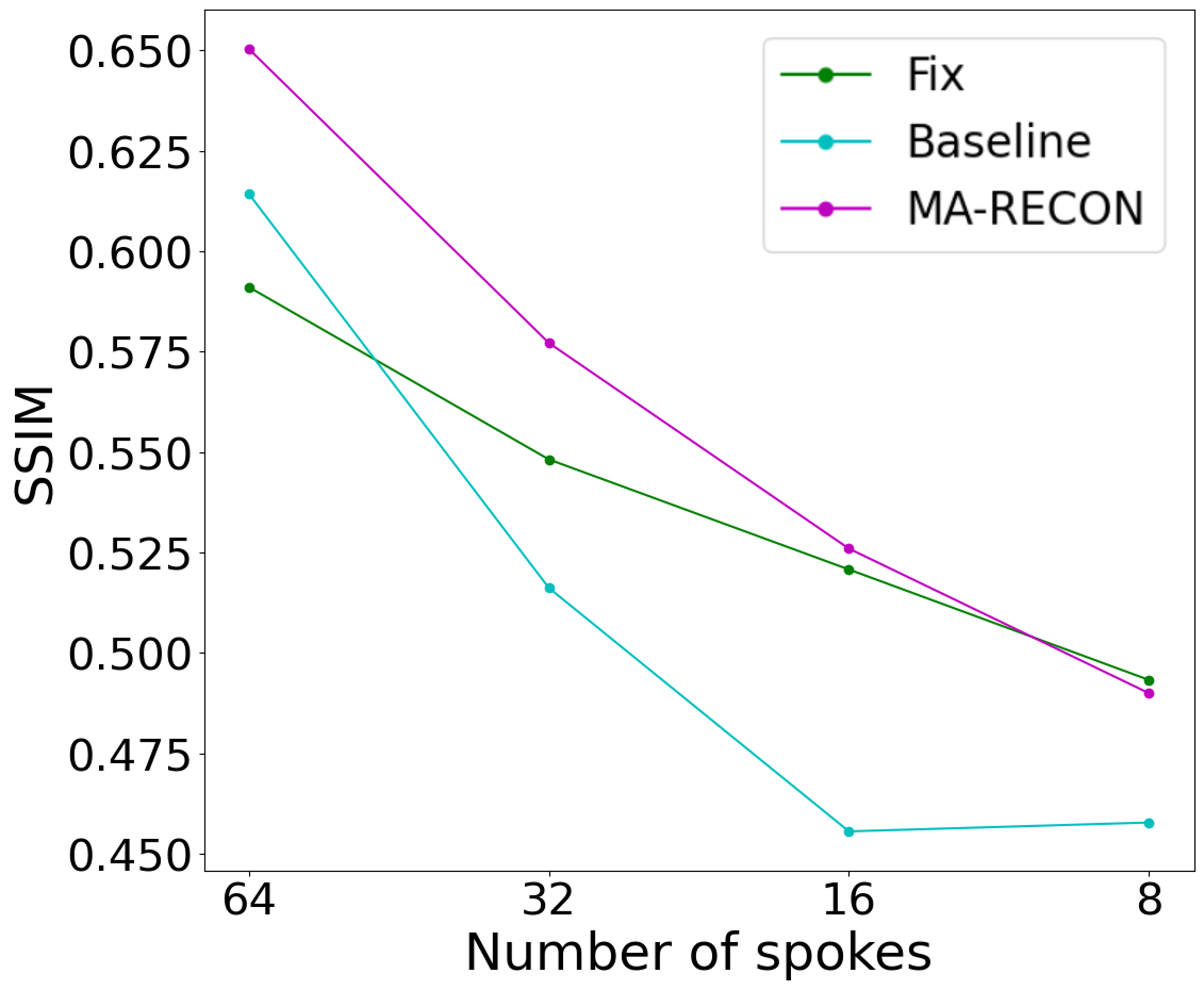}}
     \subfigure[]{\includegraphics[width=0.45\textwidth]{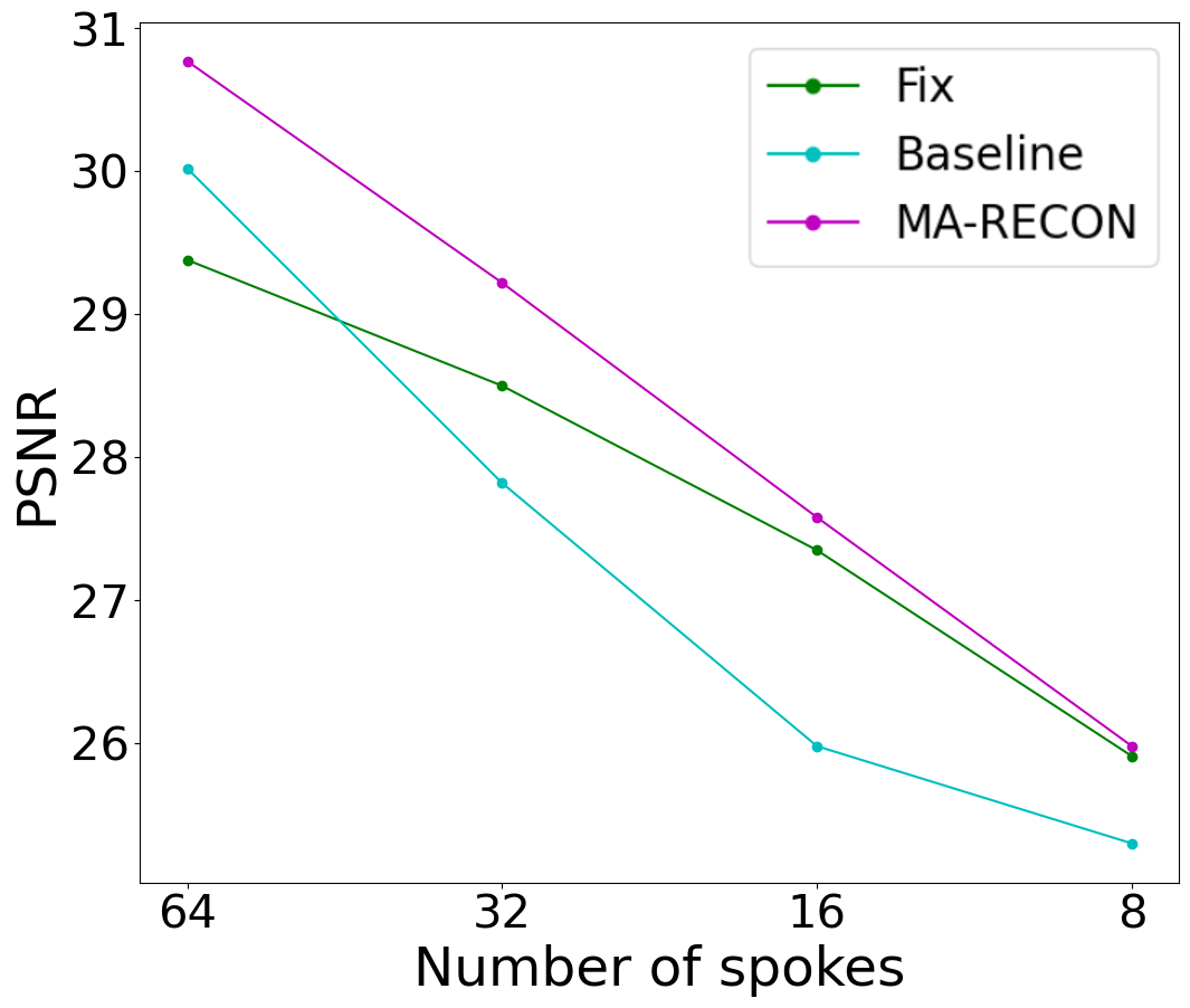}}
          
     \subfigure[]{\includegraphics[width=0.45\textwidth]{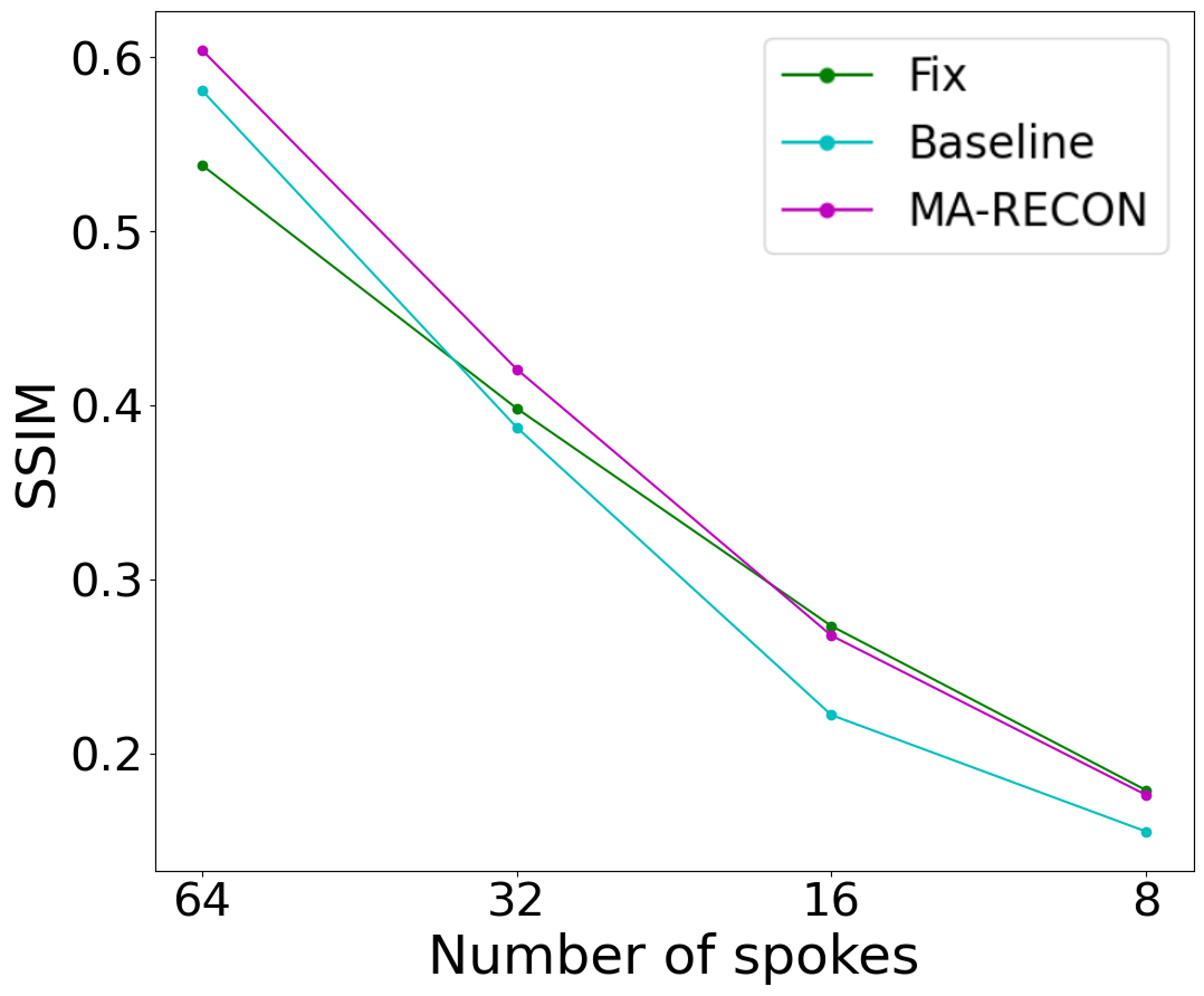}}
     \subfigure[]{\includegraphics[width=0.45\textwidth]{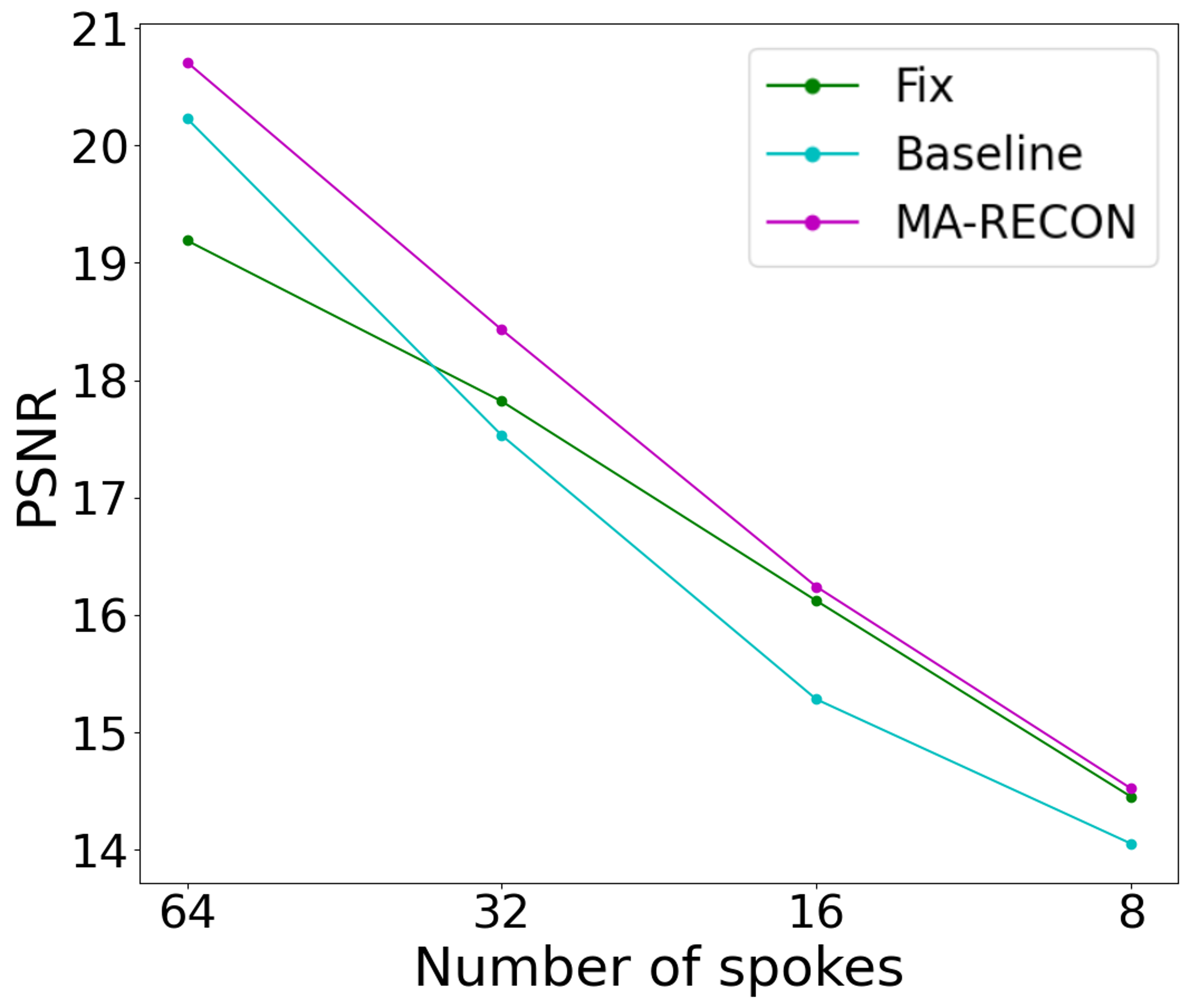}}
     
    \caption{Reconstruction accuracy for radial sampling masks with different numbers of spokes during inference. (a/b) SSIM and PSNR on all the images; (c/d) SSIM and PSNR for pathological regions}
    \label{fig:radial_results}

\end{figure}

\begin{figure}[t!]
     \centering
     \subfigure[]{\includegraphics[width=0.45\textwidth]{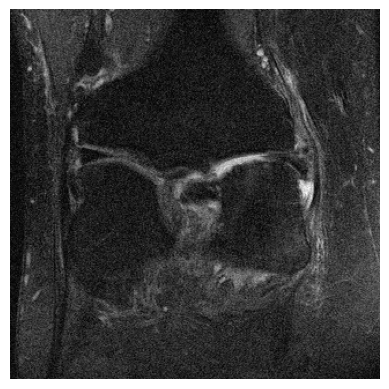}}
     \subfigure[]{\includegraphics[width=0.45\textwidth]{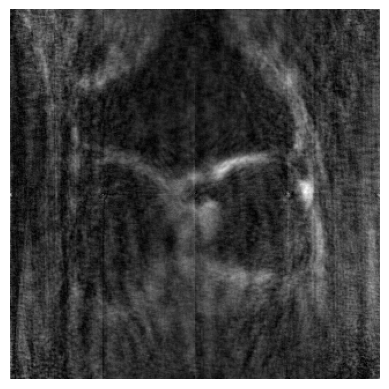}}
     
     \subfigure[]{\includegraphics[width=0.45\textwidth]{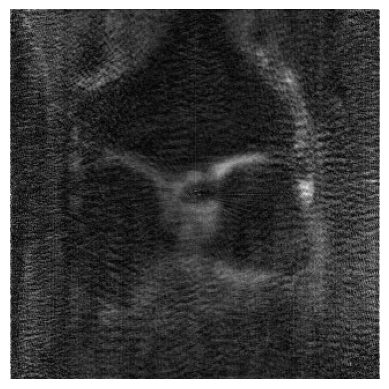}}
     \subfigure[]{\includegraphics[width=0.45\textwidth]{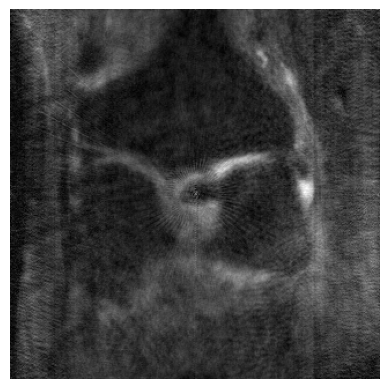}}
     
    \caption{Reconstruction for variation in the acquisition process; radial sampling mask with 32 spokes - R$\approx$10 in inference (a) target image, (b) fixed model (PSNR=29.92, SSIM=0.5849), (c) baseline model (PSNR=27.84, SSIM=0.4832), and; (d) MA-RECON (PSNR=31.24, SSIM=0.6447)}
    \label{fig:radial_recon_imgs}

\end{figure}

Fig.~\ref{fig:intro}(a-d) presents representative reconstruction results in cases of similar anatomical distribution and variation in the acquisition process, i.e. knee images with R=4 and different sampling masks in the test. Table~\ref{table:results_knee} summarizes the performance metrics of the models on the entire image and the clinically relevant regions for the same anatomical distribution (knee) but with variations in the acquisition process; different sampling masks, equispaced for training and random for testing. In all cases, the MA-RECON model has significantly better reconstruction accuracy (Paired student's t-test, p$\ll$0.01). More importantly, the improved performances are primarily evident in the clinically-relevant regions, which are critical for clinical diagnosis. Further, as can be seen in Figures~\ref{fig:radial_results} and ~\ref{fig:radial_recon_imgs}, similar results were obtained when using radial masks with different numbers of spokes during inference. The MA-RECON model also performs better for undersampled k-space that is sampled with an entirely different sampling pattern than used during training.
While our results may not reach the pinnacle of state-of-the-art performance, they effectively demonstrate the robustness of our proposed network. These results underscore the potential of our method and highlight the contribution of the mask as a prior knowledge source to the network, enhancing its capacity to improve generalization capabilities
\begin{table*}[t!]
    
    \caption{Reconstruction accuracy for variations in the acquisition process; test on random masks, and the same anatomical distribution; knee.}
    \label{table:results_knee}
    
    \centering
    \resizebox{\textwidth}{!}{
        \begin{tabular}{lllll}
        \toprule
        Image region & R test  & Model& PSNR & SSIM \\
        \midrule
        \multirow{6}{*}{Entire image} 
        &\multirow{3}{*}{4}
        &Fixed  &29.47 +/- 5.453 &	0.6374 +/- 0.2411 \\
        &&  Baseline & 30.08 +/- 5.931 & 0.6519 +/- 0.239 \\
        && \textbf{MA-RECON} &  \textbf{30.23 +/- 5.955}	& \textbf{0.6592 +/- 0.2357}\\
       \cmidrule{2-5}
        &\multirow{3}{*}{8}
         &Fixed  &27.07 +/- 4.672	& 0.5403 +/- 0.2815 \\
        &&  Baseline & 27.42 +/- 4.636	& 0.552 +/- 0.2674 \\
        && \textbf{MA-RECON} &  \textbf{27.49 +/- 4.665}	& \textbf{0.5549 +/- 0.2597}\\
       \cmidrule{1-5}
       
        \multirow{6}{*}{Pathological regions} 
         &\multirow{3}{*}{4}
        & Fixed  &	19.62 +/- 4.839&	0.5815 +/- 0.2899\\
        && Baseline &20.12 +/- 5.008&	0.6091 +/- 0.281\\
        &&  \textbf{MA-RECON} 	&\textbf{20.43 +/- 4.973}&	\textbf{0.6206 +/- 0.2713}\\ 
        \cmidrule{2-5}
        &\multirow{3}{*}{8}
        & Fixed  &	16.77 +/- 4.577&	0.391 +/- 0.3098\\
        && Baseline &17.06 +/- 4.713&	0.4096 +/- 0.3123\\
        &&  \textbf{MA-RECON} 	& \textbf{17.31 +/- 4.644}&	\textbf{0.429 +/- 0.3048}\\ 
       \bottomrule
                \end{tabular}
                }
\end{table*}    

\vspace{1cm}

\subsection{Variation in the anatomical distribution}
\begin{figure*}[t!]
     \centering
     \includegraphics[width=0.75\textwidth]{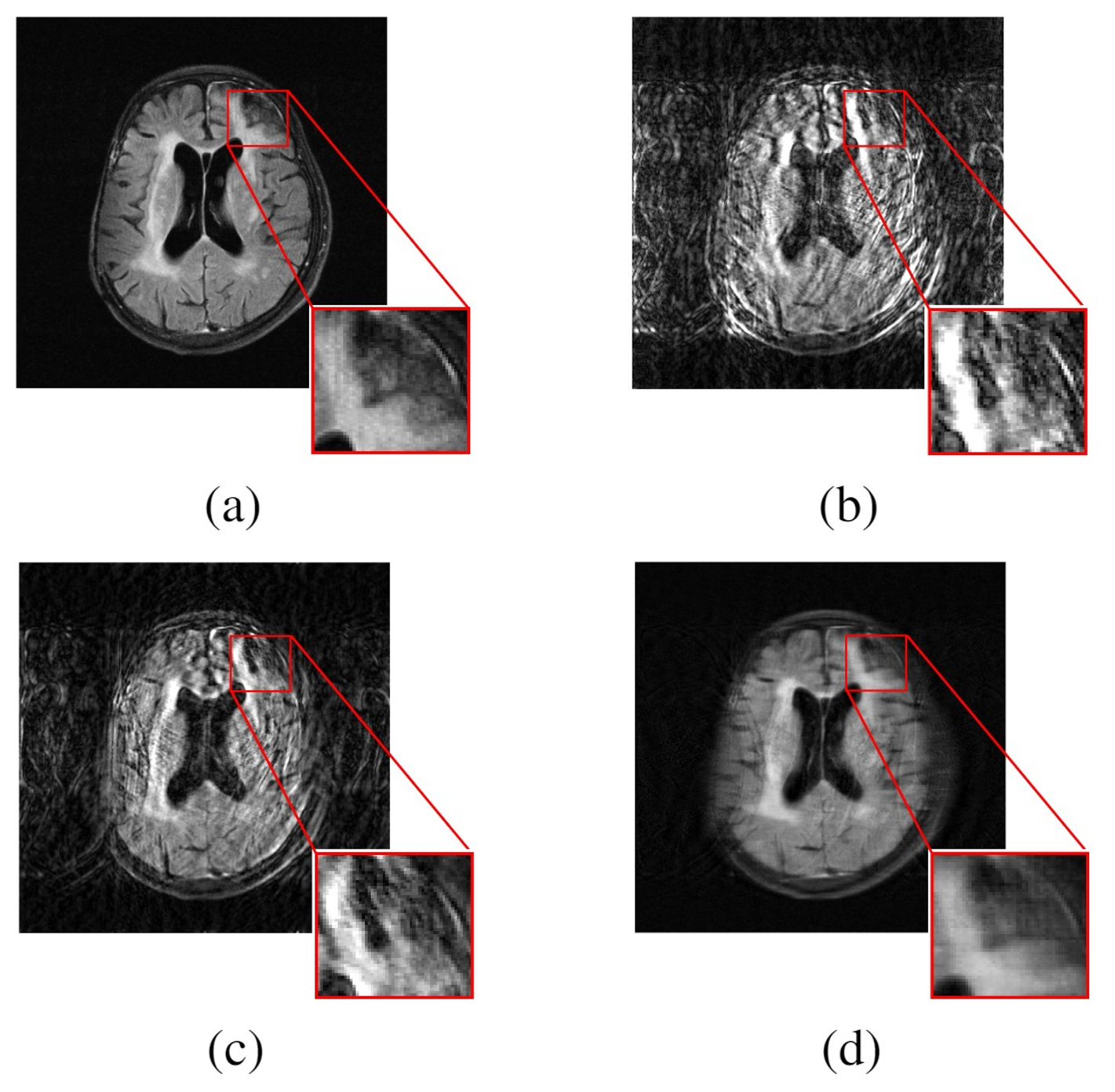}
    \caption{Reconstructed images along with zoom-in clinical annotation (red bounding box) for variation in the anatomical distribution (knee data on training, brain data on test) undersampled with equispaced mask and R=4 in test. (a) 
    target image, (b) fixed model (PSNR=14.85/10.92, SSIM=0.2405/0.364), (c) baseline model (PSNR=20.17/15.53, SSIM=0.4315/0.5377), and; (d) MA-RECON (PSNR=27.75/24.67, SSIM=0.7446/0.7779)}
    \label{fig:results_brain}
\end{figure*}

Fig.~\ref{fig:intro}(e-h) and ~\ref{fig:results_brain} present representative reconstruction results in cases of variation in the anatomical distribution, i.e. training on a knee dataset and inference on a brain dataset.  
Table~\ref{table:results_brain} summarizes the performance metrics of the models on the entire image and the clinically relevant regions when exposed to variations in the anatomical distribution. The MA-RECON performed significantly better (Paired student's t-test, p$\ll$0.01) than the models trained with and without mask augmentation. 
These results demonstrate the robustness of our mask-aware model in cases of variation in the anatomical distribution for the entire image and for the clinically-relevant regions.

\begin{table*}[t!]
    
    \caption{Reconstruction accuracy for variation in the anatomical distribution; training on knee dataset and inference on brain.}
    \label{table:results_brain}
    \centering
    \resizebox{\textwidth}{!}{
        \begin{tabular}{lllll}
        \toprule
        Image region & R test  & Model& PSNR & SSIM \\
        \midrule
        \multirow{6}{*}{Entire image} 
        &\multirow{3}{*}{4}
        &Fixed  &22.23 +/- 7.154&	0.5093 +/- 0.2493 \\
        &&  Baseline &	23.35 +/- 5.51&	0.5455 +/- 0.1786 \\
        && \textbf{MA-RECON} &  \textbf{29.92 +/- 4.55}	& \textbf{0.7442 +/- 0.139}\\
       \cmidrule{2-5}
        &\multirow{3}{*}{8}
         &Fixed  &22.48 +/- 5.041 &	0.412 +/- 0.2257 \\
        &&  Baseline & 23.53 +/- 4.57&	0.4587 +/- 0.2068 \\
        && \textbf{MA-RECON} &  \textbf{26.64 +/- 4.189}&	\textbf{0.6628 +/- 0.1576}\\
       \cmidrule{1-5}
       
        \multirow{6}{*}{Pathological regions} 
         &\multirow{3}{*}{4}
        & Fixed  &	16.41 +/- 8.643&	0.3731 +/- 0.4088\\
        && Baseline &17.65 +/- 7.585&	0.4185 +/- 0.4031\\
        &&  \textbf{MA-RECON} 	&\textbf{23.8 +/- 7.615}&	\textbf{0.6421 +/- 0.2506}\\ 
        \cmidrule{2-5}
        &\multirow{3}{*}{8}
        & Fixed  &	16.87 +/- 6.588&	0.294 +/- 0.356\\
        && Baseline &17.68 +/- 7.031&	0.3007 +/- 0.3833\\
        &&  \textbf{MA-RECON} 	& \textbf{20.78 +/- 7.396}	&\textbf{0.4544 +/- 0.2891}\\ 
       \bottomrule
        \end{tabular}
        }
\end{table*}

\subsection{Comparison with SOTA E2E VarNet}
 We performed an evaluation of the normalized mean squared error (NMSE) across a total of 55 slices, collected from five distinct subjects, with each subject contributing 11 slices. Our results indicate that our method achieved an NMSE of $0.02 \pm 0.0025$, outperforming the E2E VarNet method, which registered an NMSE of $0.029 \pm 0.0092$, as confirmed by a paired Student's t-test (p $\ll$ 0.01).
 
\begin{figure*}[t!]
     \centering
     \subfigure[]{\includegraphics[width=0.23\textwidth]{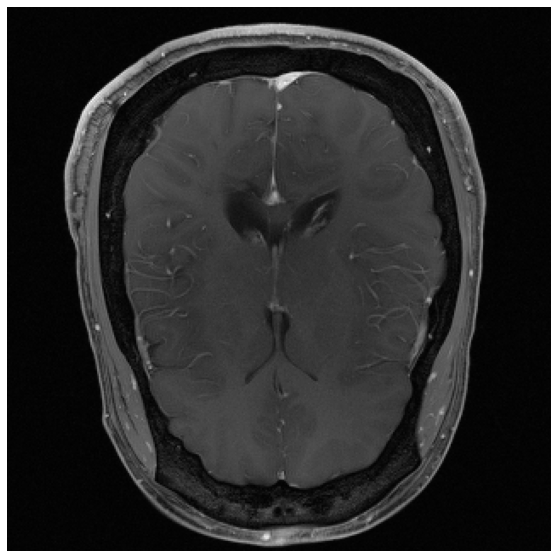}}
     \subfigure[]{\includegraphics[width=0.23\textwidth]{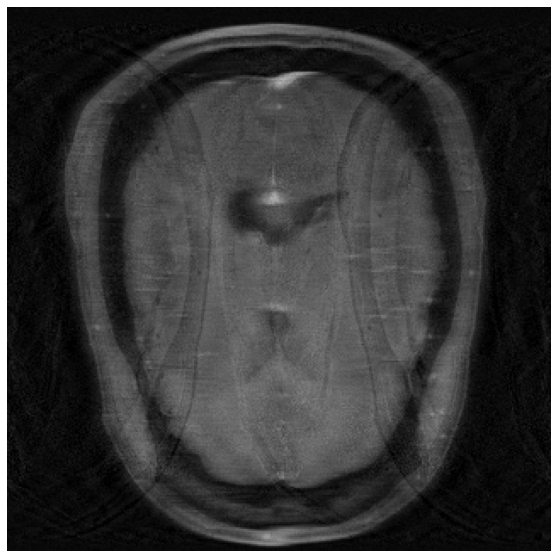}}
     \subfigure[]{\includegraphics[width=0.23\textwidth]{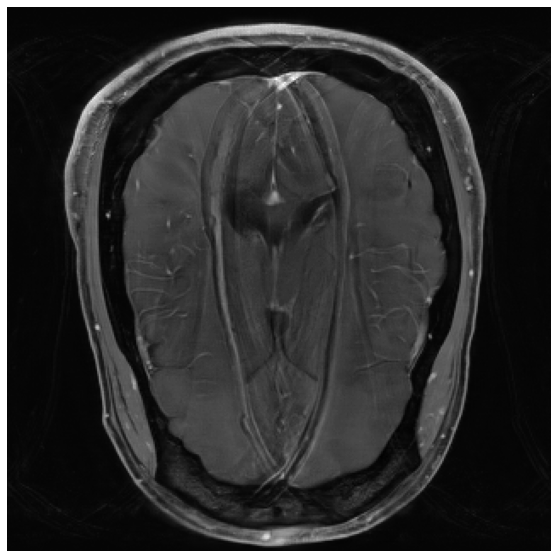}}
     \subfigure[]{\includegraphics[width=0.23\textwidth]{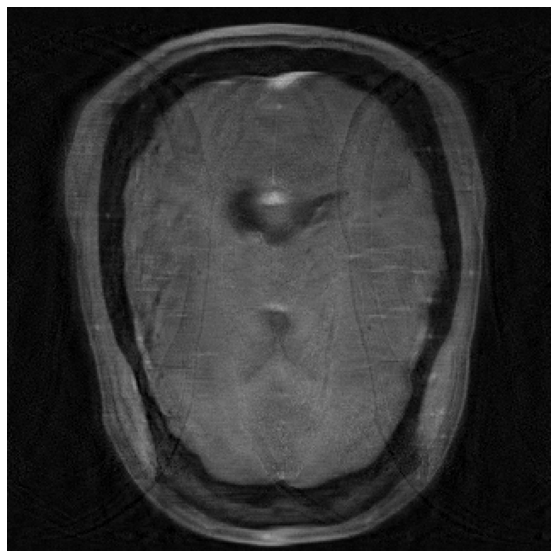}}

    \subfigure[]{\includegraphics[width=0.23\textwidth]{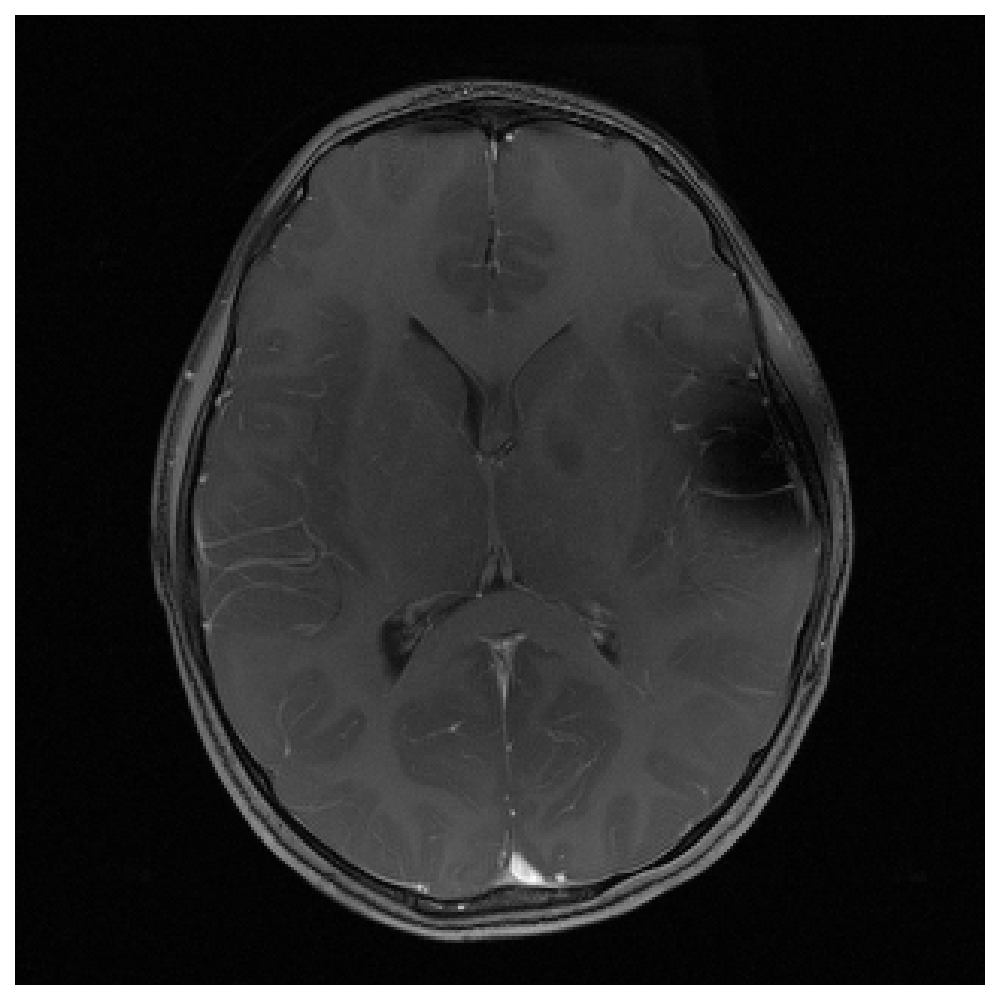}}
     \subfigure[]{\includegraphics[width=0.23\textwidth]{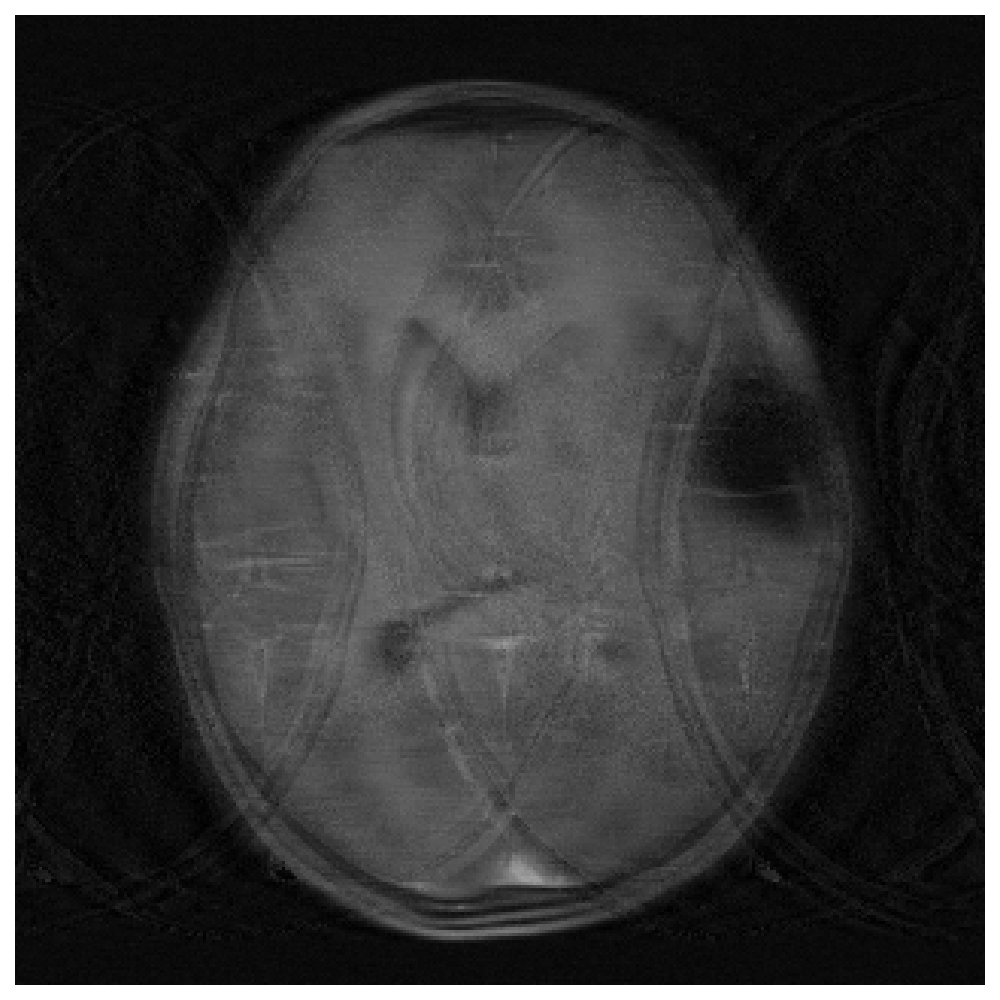}}
     \subfigure[]{\includegraphics[width=0.23\textwidth]{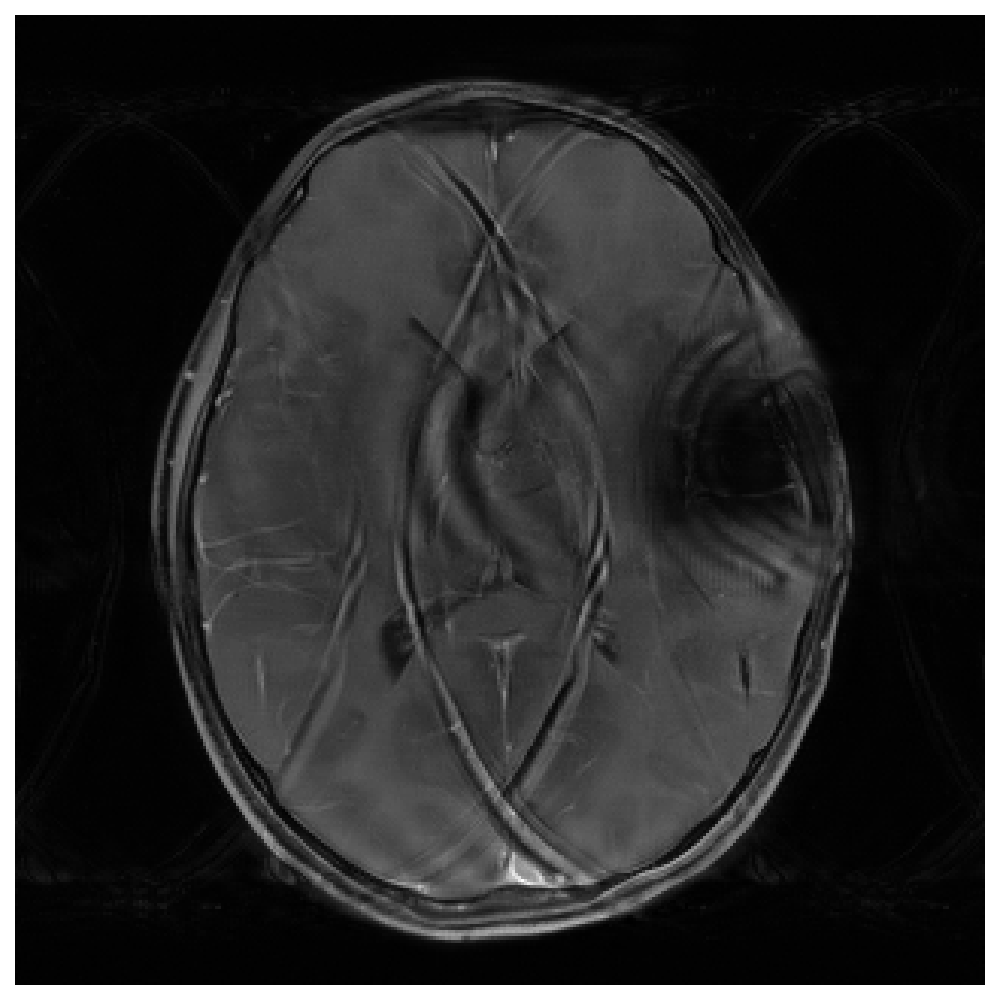}}
     \subfigure[]{\includegraphics[width=0.23\textwidth]{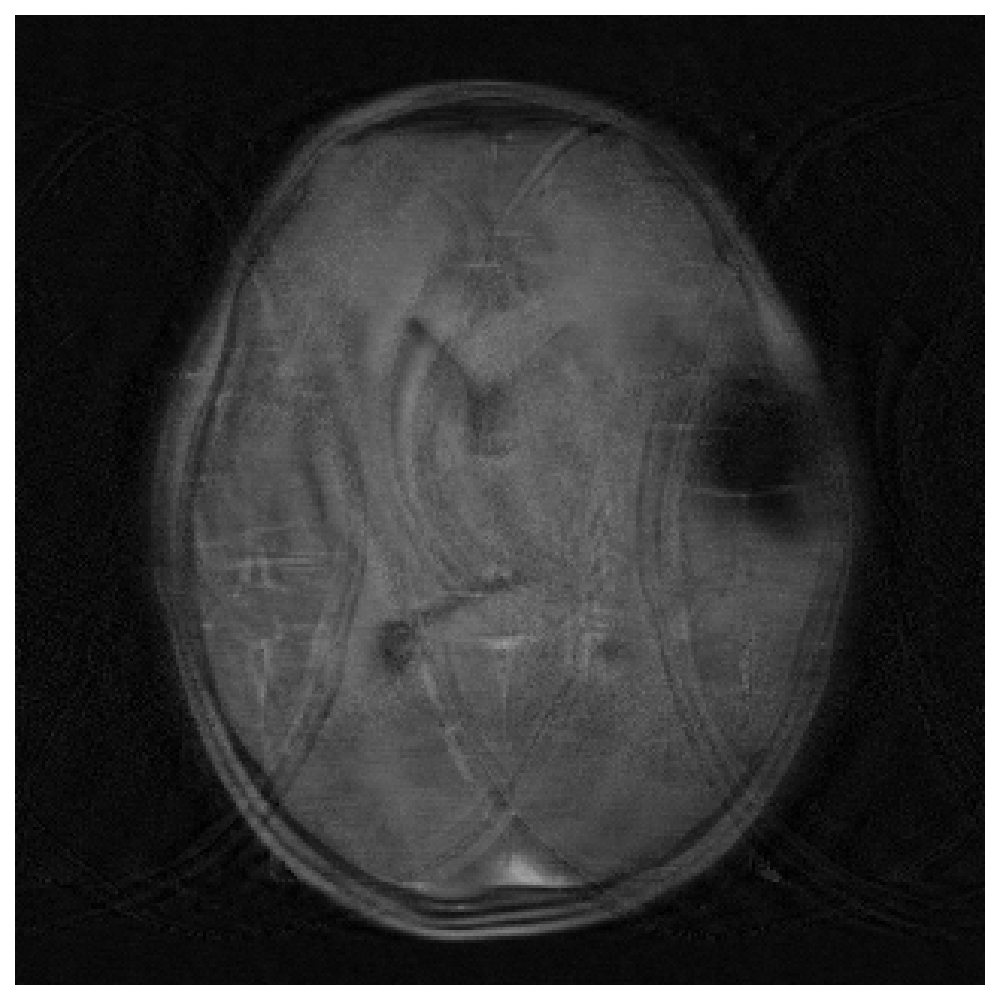}}

    \subfigure[]{\includegraphics[width=0.23\textwidth]{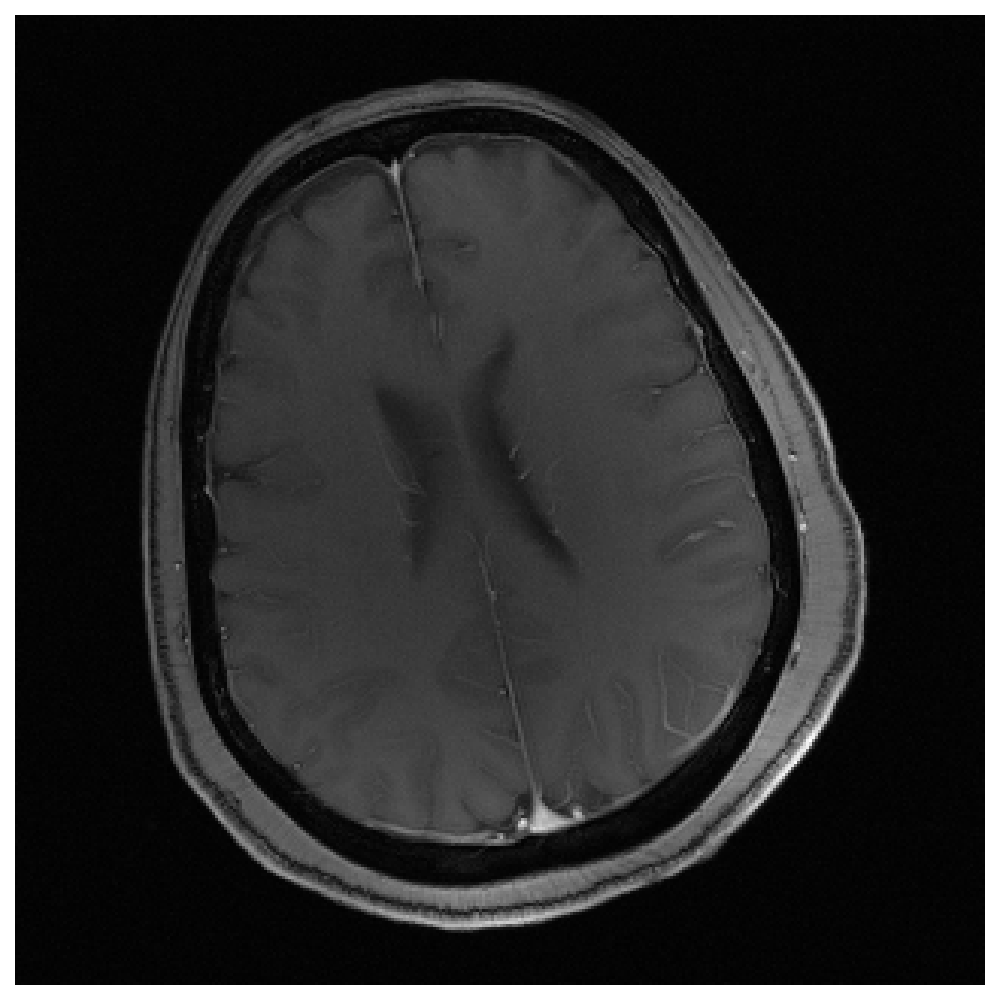}}
     \subfigure[]{\includegraphics[width=0.23\textwidth]{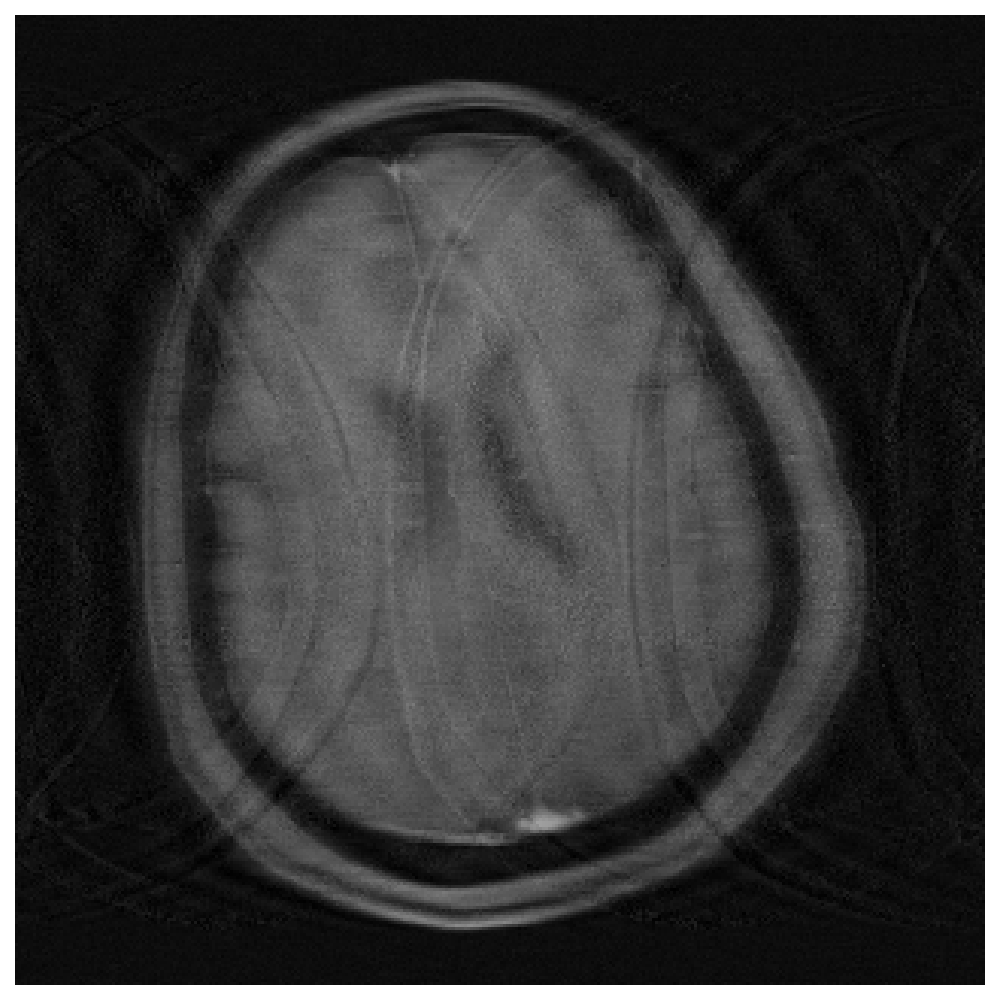}}
     \subfigure[]{\includegraphics[width=0.23\textwidth]{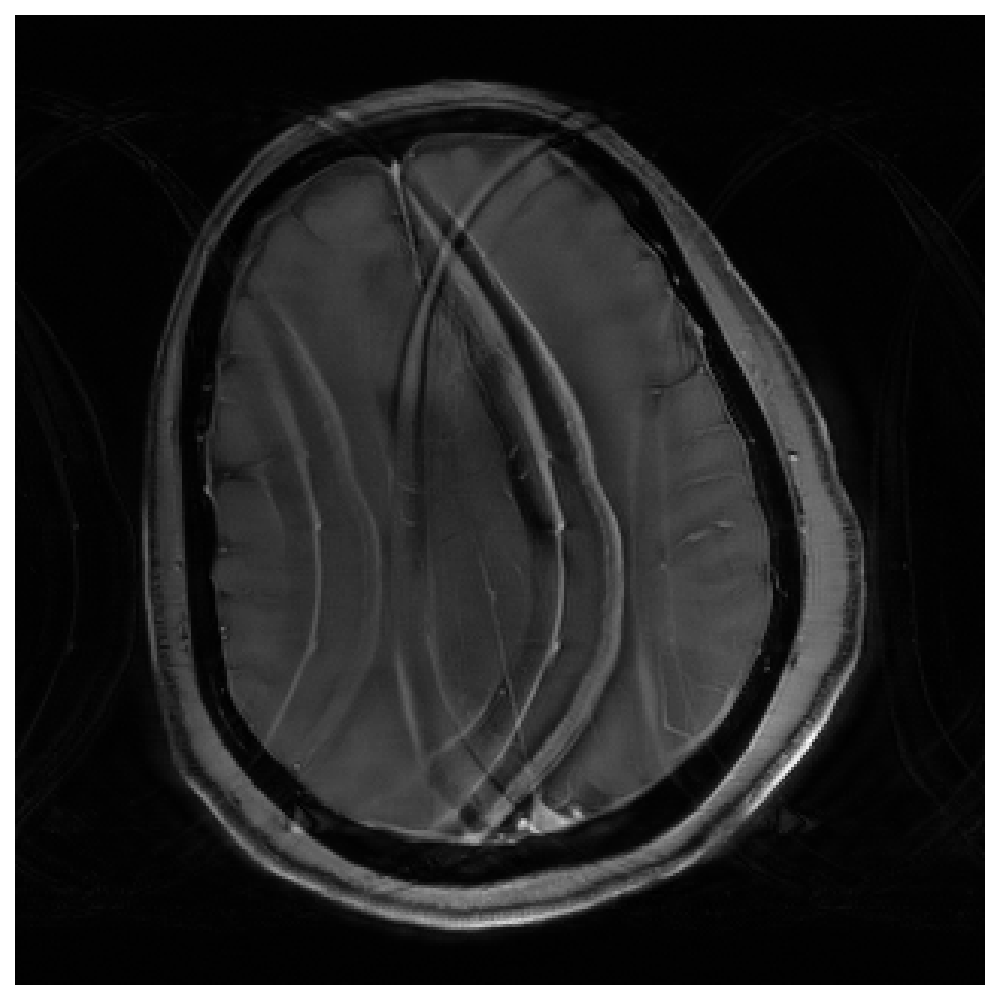}}
     \subfigure[]{\includegraphics[width=0.23\textwidth]{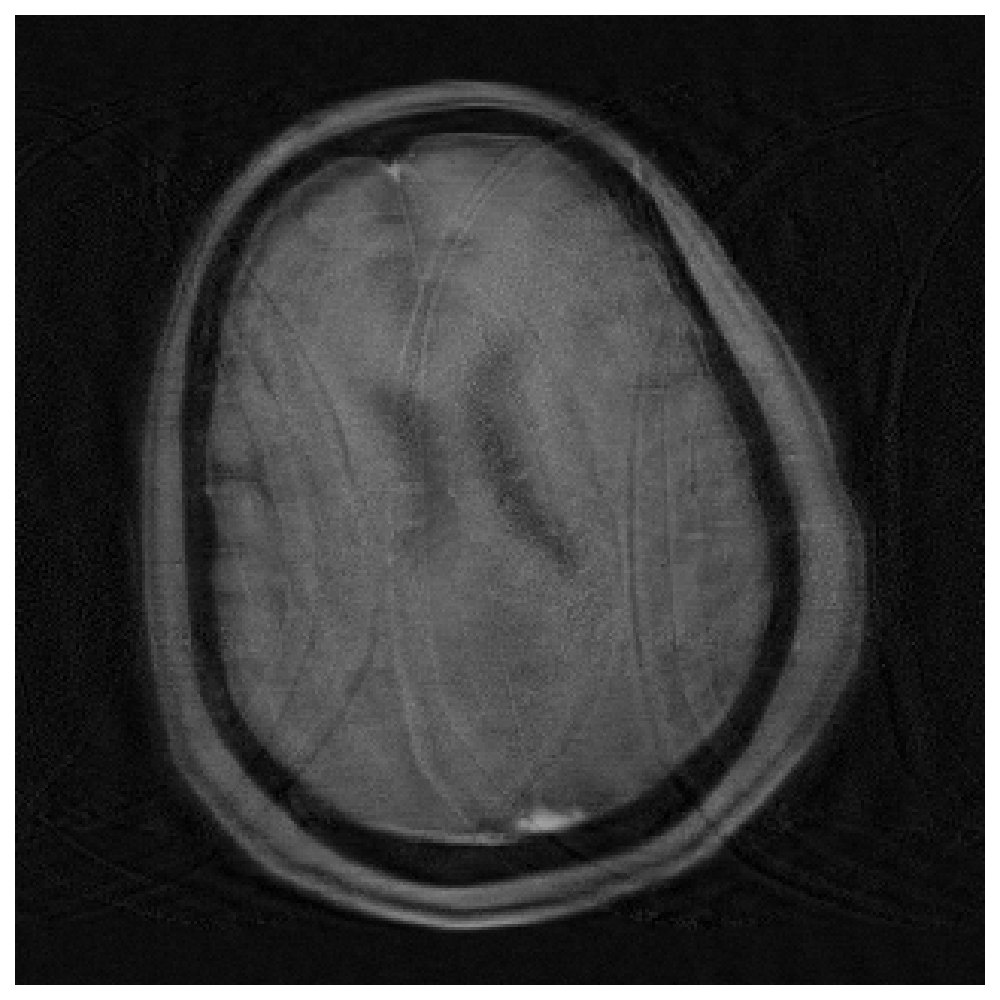}}

    \caption{Representative reconstruction results of E2E VarNet and MA-RECON against anatomical distribution shift (a/e/i) target image, (b/f/j) undersampled image, (c/g/k) E2E VarNet reconstruction, (d/h/l) MA-RECON reconstruction}
    \label{fig:e2evarnet}

\end{figure*}

 Fig.~\ref{fig:e2evarnet} presents representative reconstruction results of E2E VarNet and MA-RECON against anatomical distribution shifts. As seen in the E2E VarNet reconstruction image (c), there are prominent undersampling artifacts in the center of the image. This is in contrast to our MA-RECON model, which did not create additional artifacts and reconstructed some of them.

\section{Discussion and Conclusion}

We present MA-RECON, a mask-aware DNN architecture tailored for robust k-space interpolation. Relative to conventional methods, the historically observed stability gap in DNN-oriented methods reflects an incomplete generalization of the associated ill-posed inverse problem, effectively limiting their clinical application in MRI reconstruction from undersampled data. 
Existing methodologies strive to bridge this stability gap through data augmentation and physically inspired loss functions. Our mask-aware DNN strategy enhances the generalization capability by embedding the undersampling mask within the network infrastructure and integrating a suitable training scheme incorporating samples generated using diverse undersampling masks. 

Our investigations reveal that incorporating the undersampling mask as a component of the DNN architecture elevates the generalization capability, particularly in regions of clinical relevance, exceeding previously proposed data augmentation techniques across multiple scenarios. These include altering the undersampling mask, modifying the sampling factor, and deploying the DNN for the reconstruction of images procured from disparate anatomical regions.

Our MA-RECON model's reconstruction still needs enhancement to reach clinical-quality images with out-of-anatomical distribution data. However, the absence of additional artifacts during reconstruction suggests that MA-RECON has better generalization capabilities compared to the current SOTA method. It is pertinent to mention that the SOTA has demonstrated successful reconstruction capabilities in some cases, even in the presence of distribution shifts. However, as mentioned and described above, there have been instances where the method has failed to reconstruct, thus highlighting the importance of being aware of such limitations.

Notwithstanding, our study is not without limitations. Our evaluation was confined to single-coil volumes, and we examined our hypothesis using a specific baseline architecture and implemented only a singular method to encode the undersampling mask into the DNN architecture. While we expect our conclusions to remain valid in more clinically pertinent multi-coil environments and advanced DNN strategies, future work will need to confirm this. 
Moreover, we employed quantitative metrics such as PSNR and SSIM to objectively validate the merits of our approach. However, PSNR and SSIM may not align fully with the practical requirements of professional radiologists. Hence, an in-depth investigation is necessary before integrating these proposed techniques into clinical practice. 
Lastly, our experiments utilized the publicly available fastMRI data, which is restricted to two anatomical regions and specific MR sequences. Consequently, it may not encompass the full spectrum of possible MRI scans. The influence of our approach warrants further exploration across various MRI scans and sequences in future studies.

In summary, our mask-aware strategy holds promise for enhancing the generalization capacity and robustness of DNN-based methodologies for MRI reconstruction from undersampled k-space data. In turn, our approach offers the potential to expedite the integration of DNN-based MRI reconstruction techniques into the clinical domain.

\section*{Acknowledgements}
This work was supported in part by research grants from the Israel-US Binational Science Foundation, the Israeli Ministry of Science and Technology, the Israel Innovation Authority, and the joint Microsoft Education and the Israel Inter-university Computation Center (IUCC) program. N.A. is sponsored in part by a PhD scholarship for outstanding students in Data Sciences from the Israel Council for Higher Education.

\section*{Declaration of Generative AI}
During the preparation of this work, the author(s) used ChatGPT in order to improve readability. After using this tool/service, the author(s) reviewed and edited the content as needed and take(s) full responsibility for the content of the publication.

\section*{Appendix}

\begin{figure*}[h]
    \centering
    \includegraphics[width=\textwidth]{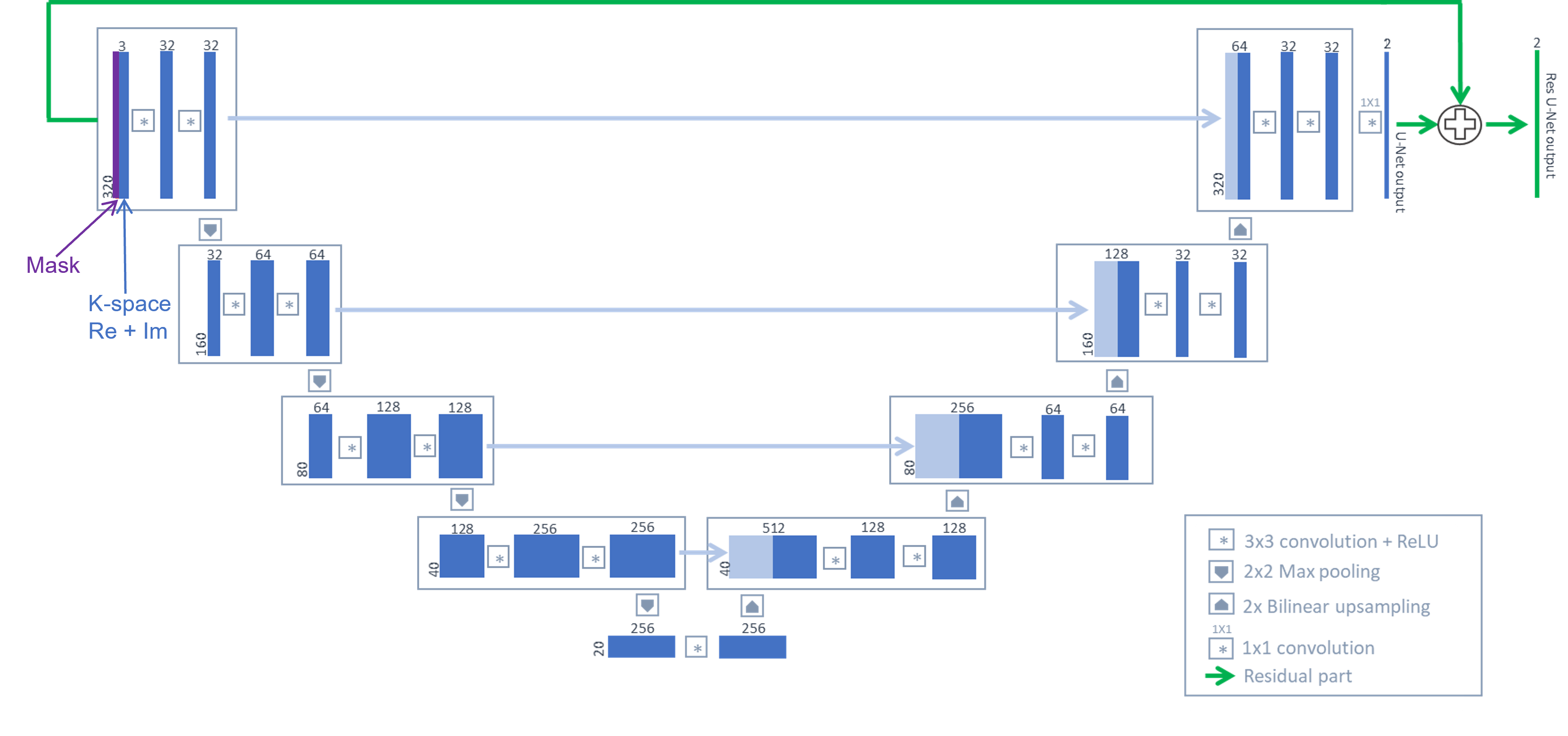}
    \label{fig:architecture}
    \caption{Model architecture. The network input is 3-channel data representing the concatenation of complex k-space data with the mask. The output is a summation of the network and the 2-channel k-space input.}

\end{figure*}

\bibliographystyle{model2-names.bst}\biboptions{numbers}

\bibliography{main}

\end{document}